\documentclass{ws-mpla}
\usepackage[utf8]{inputenc}
\usepackage[T1]{fontenc}
\usepackage{graphicx}
\usepackage[super]{cite}
\usepackage{amsmath}
\usepackage{amssymb}
\usepackage{capt-of}
\usepackage{hyperref}
\usepackage{subfigure}
\usepackage{booktabs,array}
\usepackage[dvipsnames]{xcolor}
\begin{document}
\title{Phenomenological implications of two simple modifications to Tri-Bimaximal mixing}
\author{Kanwaljeet S. Channey}
\address{Department of Physics and Astrophysics, University of Delhi,\\
Delhi, 110007,
India.\\
kjschanney@outlook.com}
\author{Sanjeev Kumar}
\address{Department of Physics and Astrophysics, University of Delhi,\\
Delhi, 110007,
India.\\
sanjeevkumarverma@outlook.in}
\maketitle
\begin{abstract}

We study the phenomenological implications of breaking the Tri-Bimaximal (\text{TBM}) mixing
in such a way that either first or second column of \text{TBM} mixing matrix remains invariant. We
present two such textures and confront them with the experimental data. We give the
predictions of these textures for atmospheric mixing angle $\theta_{23}$ and
Dirac-type CP violating phase $\delta$ that will be measured
in the future experiments.

\end{abstract}

One of the unresolved mysteries of the neutrino physics is the origin of the
neutrino masses. If the three neutrino masses are distinct and non zero, the mass and flavour eigen states
of neutrinos are not identical. The flavour eigen states ($\nu_f$) can be written
as a linear combination of the mass eigen states ($\nu_i$): 

\begin{equation}
  \label{eq:mixing}
  \nu_{f}=U \nu_{i}.
\end{equation}
Here, $i = 1,2,3$ and $f=e, \mu, \tau$. The neutrino mixing
matrix, $U$, is given as \cite{Olive2014}

\begin{equation}
 \label{eq:pmns}
U = \left(\begin{array}{ccc}
                   c_{12} c_{13} & s_{12} c_{13} & s_{13} e^{-\iota \delta} \\
                   -s_{12}c_{23}-c_{12} s_{23} s_{13} e^{\iota \delta} & c_{12} c_{23}-s_{12} s_{23} s_{13} e^{\iota \delta} & s_{23} c_{13} \\
                   s_{12} s_{23}-c_{12} c_{23} s_{13} e^{\iota \delta} & -c_{12} s_{23}-s_{12} c_{23} s_{13} e^{\iota \delta} & c_{23} c_{13} \\
                 \end{array}
               \right),
             \end{equation}
             where $s_{ij}=\sin \theta_{ij}$ and $c_{ij}=\cos \theta_{ij}$ for
             $i,j=1,2,3$ and $\theta_{12}$, $\theta_{23}$, $\theta_{13}$ are the
             three mixing angles. The phase $\delta$ is responsible for
             Dirac-type CP violation in neutrino oscillations.
The mixing matrix (U) diagonalizes the neutrino mixing matrix ($M_{\nu}$) as

\begin{equation}
  \label{eq:diag}
U^T M_{\nu} U = M^{\text{diag}}_{\nu}
\end{equation}
where, the diagonal neutrino mass matrix is $M^{\text{diag}}_{\nu} = \text{diag} \{m_1,m_2e^{2\iota \alpha}, m_3 e^{2
  \iota \beta} \}$.
Here, the phases $\alpha$ and $\beta$ are the Majorana phases. These phases do not appear in the oscillation probabilities. However, they can
be constrained in the neutrino less double beta decay experiments. If the three
neutrino masses ($m_1, m_2$ and $m_3$), three neutrino mixing angles
($\theta_{12}, \theta_{23}$ and $\theta_{13}$) and the three CP violating phases
($\alpha, \beta, \delta$) are known, the neutrino mass matrix can be
reconstructed as

\begin{equation}
  \label{eq:dia}
  M_{\nu} = U^* M^{\text{diag}}_{\nu} U^{\dagger}.
\end{equation}

The reconstruction of the neutrino mass matrix results into many possible
structures. The experimental measurements of a non-zero $\theta_{13}$ \cite{T2K_NON_ZERO_Theta13_PhysRevLett.107.041801,Daya_Bay_reactor_angle_PhysRevLett.108.171803,RENO_THETA13_PhysRevLett.108.191802,Double_CHOOZ_reactor_angle_PhysRevLett.108.131801} ruled out
many such structures. One such structure corresponded to the Tri-BiMaximal (TBM)
mixing \cite{Harrison2002167} that predicts $\theta_{13} = 0$ and $\theta_{23} = \frac{\pi}{4}$. After
the measurement of non-zero $\theta_{13}$, the TBM scheme cannot be compatible with
the neutrino data at the leading order. Yet, one can modify the mass matrix
corresponding to the TBM mixing ($M_{\text{TBM}}$) by adding some correction terms that
break the underlying symmetry of $M_{\text{TBM}}$. However, such modifications need not
break the symmetry of $M_{\text{TBM}}$ completely. We can modify $M_{\text{TBM}}$ in such a
manner that the resulting mixing matrix still has its first or second column
identical to the TBM mixing matrix $U_{\text{TBM}}$. Such mixing schemes can be called
Tri-Maximal (TM) mixing of first and second kind ($\text{TM}_1$
\cite{Xing2007a,Lam2006a,Albright2008,Albright2010,He2011a,Kumar2013} and $\text{TM}_2$ \cite{Xing2007a,Harrison2004,Friedberg:2006it,Xing2006,Bjorken2006,He2011a,Albright2008,Albright2010,Kumar2010a,Kumar2013}),
respectively.

In the present work, we propose two simple textures that can modify $M_{\text{TBM}}$ to
have non-zero $\theta_{13}$ and non-maximal $\theta_{23}$ while preserving its
first or second eigen vector at its TBM value. We study the phenomenology of
these textures and confront them with the experimental data. Our textures are
testable at the future neutrino experiments like NO$\nu$A\cite{NOVA_2016_delta_PhysRevLett.116.151806} and T2K\cite{T2K_NON_ZERO_Theta13_PhysRevLett.107.041801} that aim to
measure the octant of $\theta_{23}$ and CP violating phase $\delta$. We also
present the predictions of these textures for the Majorana phases ($\alpha$ and
$\beta$) and the neutrino masses as measured in beta decay\cite{Drexlin2013}, neutrino less double
beta decay \cite{Rodejohann2012a} and cosmological experiments\cite{Couchot2017}.

The TBM mass matrix can be written as

\begin{equation}
  \label{eq:mtbm}
M_{\text{TBM}} = \left(
\begin{array}{ccc}
 a & c & c \\
 c & a+b & c-b \\
 c & c-b & a+b \\
\end{array}
\right).
\end{equation}
It can be diagonalized as
\begin{equation}
  \label{eq:diag}
U^T_{\text{TBM}} M_{\text{TBM}} U_{\text{TBM}} = M^{\text{diag}}_{\nu}
\end{equation}
where
  \begin{equation}
  \label{eq:utbm}
  U_{\text{\text{TBM}}}= \left( \begin{array}{ccc}
           \sqrt{\frac{2}{3}} & \sqrt{\frac{1}{3}} & 0 \\
           -\sqrt{\frac{1}{6}} & \sqrt{\frac{1}{3}} & \sqrt{\frac{1}{2}} \\
           -\sqrt{\frac{1}{6}} & \sqrt{\frac{1}{3}} & -\sqrt{\frac{1}{2}}
         \end{array} \right).
     \end{equation}
     The neutrino mass matrices for the mixing schemes TM$_1$
     \cite{Xing2007a,Lam2006a,Albright2008,Albright2010,He2011a,Kumar2013} and TM$_2$ \cite{Xing2007a,Harrison2004,Friedberg:2006it,Xing2006,Bjorken2006,He2011a,Albright2008,Albright2010,Kumar2010a,Kumar2013} can be
     written as
     \begin{equation}
       \label{eq:break}
M_{\text{TM}_{i}} = M_{\text{TBM}} + M'_{\text{TM}_i}
\end{equation}
where $i = 1, 2$ corresponds to TM$_1$ and TM$_2$ mixings. The general forms of
these modified mass matrices can be written as

\begin{equation}
  \label{eq:mtm1}
M_{\text{TM}_{1}} = \left(
\begin{array}{ccc}
 a & 2 c-d & d \\
 2 c-d & a+b+4 (c-d) & -b-c+2 d \\
 d & -b-c+2 d & a+b \\
\end{array}
\right)
\end{equation}
and
\begin{equation}
  \label{eq:mtm2}
M_{{\text{TM}}_2} = \left(
\begin{array}{ccc}
 a & c & d \\
 c & a+b-c+d & c-b \\
 d & c-b & a+b \\
\end{array}
\right).
\end{equation}
The most general form of $M'_{\text{TM}_{i}}$ will be similar to $M_{\text{TM}_i}$. However,
we do not work with the most general forms as they have too many free
parameters. We assume simple textures of $M_{\text{TBM}}$ and $M'_{\text{TM}_i}$ that are
compatible with the neutrino data. The two textures studied in this work are

\begin{equation}
  \label{eq:main}
M_i = M_0 + M'_i
\end{equation}
where $M_i =
M_{\text{TM}_{i}}$ with $c=a$ and $d = a + \mu$ for $i=1,2$. Here, the parameter
$a$, $b$, and $c$ are real while $\mu=z e^{\iota \chi}$ is a complex parameter.
Therefore, our textures
have four real parameters \textit{i.e.} $a$, $b$, $z$, and $\chi$. The
textures for $M_0$ and $M'_i$ are

\begin{equation}
  \label{eq:msses}
M_0 = \left(
\begin{array}{ccc}
 a & a & a \\
 a & a+b & a-b \\
 a & a-b & a+b \\
\end{array}
\right), M'_1 = \left(
\begin{array}{ccc}
 0 & -\mu  & \mu  \\
 -\mu  & -4 \mu  & 2 \mu  \\
 \mu  & 2 \mu  & 0 \\
\end{array}
\right), M'_{2} = \left( \begin{array}{ccc}
                     0 & 0 & \mu \\
                     0 & \mu & 0 \\
   \mu & 0 &  0\\ 
                   \end{array}
 \right).
\end{equation}
The assumption $c=a$ leads to a vanishing lowest mass eigenvalue of $M_0$. The
modification term $M'_i$ may modify the eigenvalues of $M_0$ and we may end up
with a non-zero value of
$m_1$. The parameter $z$ acts as the modification parameter and gives us a
non-zero $\theta_{13}$ and non-maximal $\theta_{23}$. The resulting textures
$M_i$ will have normal hierarchy. However, we could construct similar textures
for inverted hierarchy by setting $c=a+2b$ and $d=a+\mu$ in Eqs.
(\ref{eq:mtm1}-\ref{eq:mtm2}).

The mass matrices $M_{i}$ can be diagonalized using the relation
\begin{equation}
  \label{eq:diag}
 M^{\text{diag}}_{i} = U^T_{{\text{TM}}_i} M{_i} U_{\text{TM}_{i}}
\end{equation}
where
\begin{equation}
  \label{eq:tm1}
U_{\text{TM}_1} = \left(
    \begin{array}{ccc}
      \sqrt{\frac{2}{3}} & \frac{\cos \theta}{\sqrt{3}} & \frac{\sin \theta
                                                          }{\sqrt{3}} \\
      -\frac{1}{\sqrt{6}} &
                            \frac{\sqrt{\frac{2}{3}} \cos \theta
                            -e^{i \phi } \sin  \theta
                            }{\sqrt{2}} & \frac{e^{i \phi } \cos
                                          \theta+\sqrt{\frac{2}{3}} \sin
                                          \theta}{\sqrt{2}} \\
      -\frac{1}{\sqrt{6}} &
                            \frac{\sqrt{\frac{2}{3}} \cos \theta
                            +e^{i \phi } \sin \theta
                            }{\sqrt{2}} &
                                          \frac{\sqrt{\frac{2}{3}} \sin \theta
                                          -e^{i \phi } \cos \theta
                                          }{\sqrt{2}}
    \end{array}
  \right)
\end{equation}
and
\begin{equation}
  \label{eq:tm2}
  U_{{\text{TM}}_2} =\left(
    \begin{array}{ccc}
      \sqrt{\frac{2}{3}}
      \cos  \theta   &  \frac{1}{\sqrt{3}}  & \sqrt{\frac{2}{3}}
                                              \sin  \theta   \\
      \frac{e^{i \phi } \sin  \theta  -\frac{\cos  \theta }{\sqrt{3}}}{\sqrt{2}}&  \frac{1}{\sqrt{3}}  & \frac{-e^{i
                                                                                                         \phi } \cos  \theta  -\frac{\sin
                                                                                                         \theta  }{\sqrt{3}}}{\sqrt{2}} \\
      \frac{-\frac{\cos
      \theta  }{\sqrt{3}}-e^{i \phi } \sin
      \theta  }{\sqrt{2}}&                 \frac{1}{\sqrt{3}}   & \frac{e^{i
                                                                  \phi } \cos  \theta  -\frac{\sin
                                                           \theta  }{\sqrt{3}}}{\sqrt{2}} \\
    \end{array}
  \right).
\end{equation}
Here, $\theta$ and $\phi$ are the two free parameters of the TM$_1$ and TM$_2$
mixing matrices. The experimental values of the three mixing angles and the
Jarlskog rephasing invarient measure of CP violation $J$ \cite{Jarlskog1985} can be calculated in
terms of $\theta$ and $\phi$ using the relations

\begin{eqnarray}
  \label{eq:genform}
  \sin^2 \theta_{12} &=& \frac{|(U_i)_{12}|^2}{1-|(U_i)_{13}|^2}, \\ 
  \sin^2 \theta_{23} &=& \frac{|(U_i)_{23}|^2}{1-|(U_i)_{13}|^2},\\ \label{eq:genform2}
    \sin^2 _{13} &=& |(U_i)_{13}|^2,\label{eq:genform3}
\end{eqnarray}

and
\begin{equation}
  \label{eq:jarlskog}
J = Im[(U_i)_{11}(U_i)^*_{12}(U_i)^*_{21}(U_i)_{22}].
\end{equation}

We calculate the $M^{\text{diag}}_i$ for the textures $M_1$ and $M_2$ using Eq.
(\ref{eq:diag}). We find that the off-diagonal elements $\left( M^{\text{diag}}_1 \right)_{12},\left(
  M^{\text{diag}}_1 \right)_{13}, \left(
  M^{\text{diag}}_2 \right)_{12}$ and $\left( M^{\text{diag}}_2 \right)_{23}$
are zero identically. The remaining off-diagonal
elements, $\left(
  M^{\text{diag}}_1 \right)_{23}$ and $\left(
  M^{\text{diag}}_2 \right)_{13}$, are in general functions of $\theta$ and $\phi$. Equating them to zero,
we obtain the predictions of our textures for $\theta$ and $\phi$ in terms of
the four free parameters $a$, $b$, $z$, and $\chi$. Solving the complex equation
$(M^{\text{diag}}_1)_{23}=0$, we find that
\begin{equation}
  \label{eq:phifortm1}
\tan \phi = \frac{(2 b-3 a) \tan \chi }{3 a+2 b-4 z \sec \chi }
\end{equation}
and
\begin{equation}
  \label{eq:thetafortm1}
\tan 2 \theta = \frac{2 \sqrt{6} z \cos (\chi +\phi )}{3 a-2 b \cos 2 \phi +4 z \cos (\chi +2 \phi )}
\end{equation}
for the texture $M_1$. Similarly, equating $(M^{\text{diag}}_2)_{13}=0$, we find
that
\begin{equation}
  \label{eq:phifortm2}
 \phi = -\chi 
\end{equation}
and
\begin{equation}
  \label{eq:thetafortm2}
\tan 2 \theta = \frac{\sqrt{3} z \sin (\phi -\chi )}{ 2 b \sin 2 \phi +z \cos \phi  \sin (\phi -\chi )}
\end{equation}
for the texture $M_2$.
The predictions for the mixing angles $\theta_{12}$ and $\theta_{13}$ are
\begin{equation}
  \label{eq:mixing angles 12 and 13 for tm1}
  \sin^2 \theta_{12} = 1-\frac{4}{\cos 2 \theta +5} \text{ \hspace{0.3cm}and\hspace{0.3cm} } \sin^2 \theta_{13} = \frac{\sin ^2 \theta}{3}
\end{equation}
for the texture $M_1$ and
\begin{equation}
  \label{eq:mixing angles 12 and 13 for tm2}
\sin^2 \theta_{12} =\frac{1}{3-2 \sin ^2 \theta} \text{ \hspace{0.3cm}and\hspace{0.3cm} } \sin^2 \theta_{13} = \frac{2 \sin ^2\theta }{3}
\end{equation}
for the textures $M_2$. Here, the variables $\theta$ and $\phi$ are to be
substituted from Eqs. (\ref{eq:phifortm1}-\ref{eq:thetafortm1}) for the texture
$M_1$ and from
Eqs. (\ref{eq:phifortm2}-\ref{eq:thetafortm2}) for the texture $M_2$.
\begin{figure}[t]
  \centering
 \includegraphics[width=6.29cm]{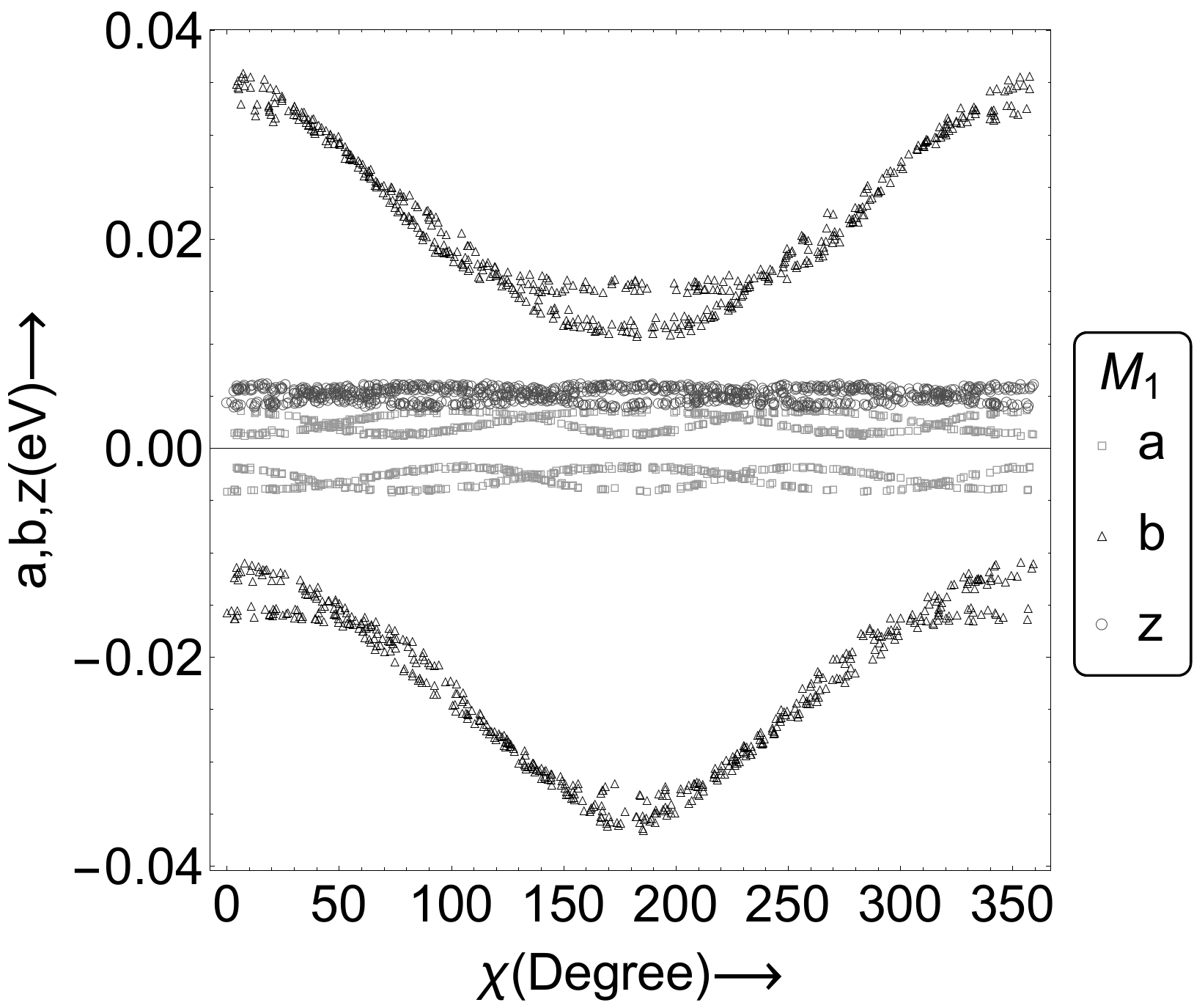}  \includegraphics[width=6.29cm]{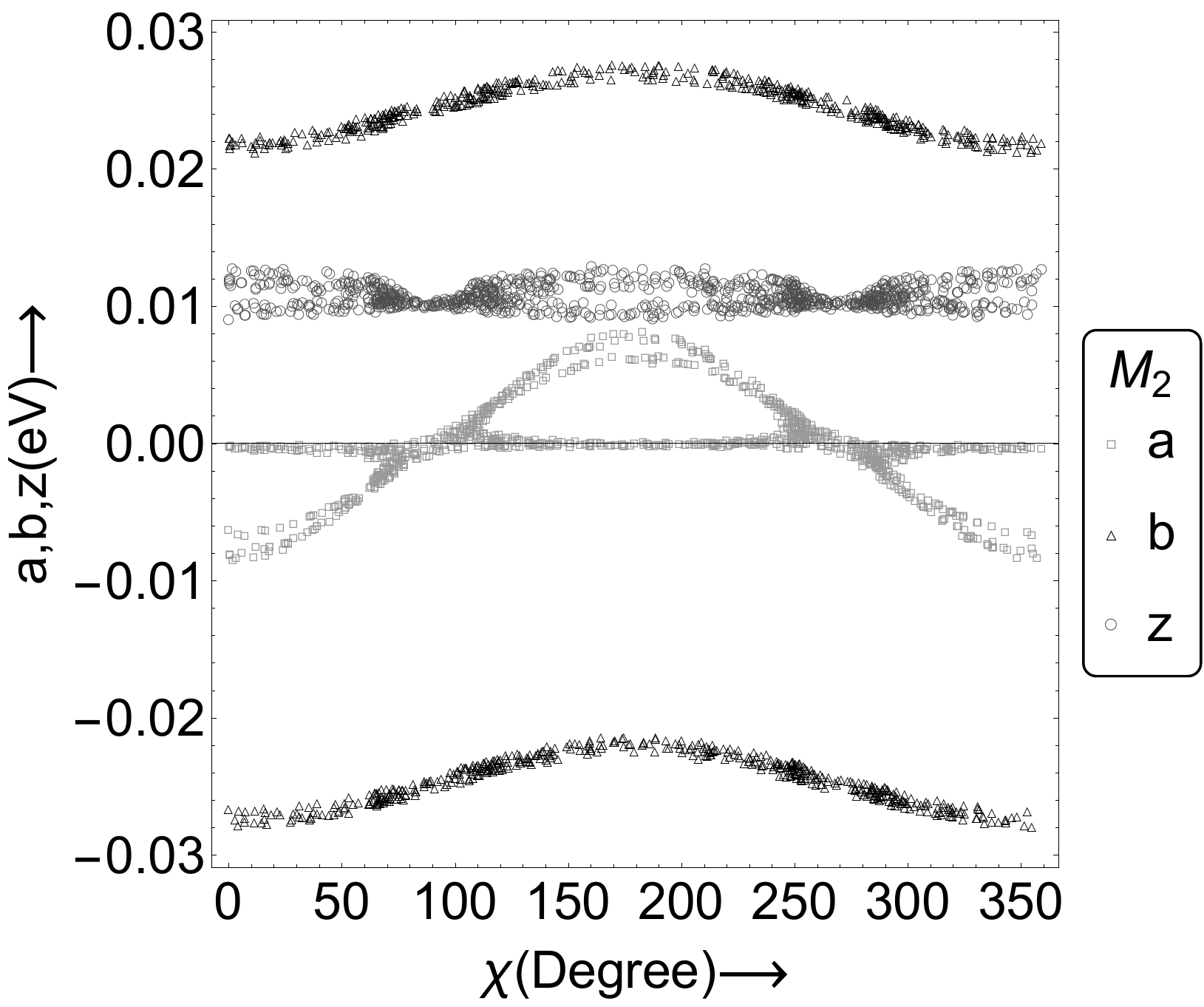}
    \caption{The allowed parameter space for $a, b, z,$ and $\chi$ for the textures M$_1$
      and M$_2$.}
 \label{massvschi}
\end{figure}

The diagonal enteries of $M^{\text{diag}}_1$ are given as
\begin{eqnarray}
  \label{eq:diag eneteries of tm1 mass matrices-11}
  \left( M^{\text{diag}}_{1} \right)_{11} &=& 0, \\
  \left( M^{\text{diag}}_{1} \right)_{22} &=& 3 a \cos ^2 \theta +2 e^{2 i \phi
                                              } \sin ^2 \theta  \left(b-2 e^{i \chi } z\right)+\sqrt{6} z \sin 2 \theta  e^{i (\chi +\phi )},  \label{eq:diag eneteries of tm1 mass matrices-22}
\end{eqnarray}
and
\begin{equation}
 \label{eq:diag eneteries of tm1 mass matrices-33}
  \left( M^{\text{diag}}_{1} \right)_{33} = 3 a \sin ^2 \theta +2 e^{2 i \phi } \cos ^2 \theta \left(b-2 e^{i \chi } z\right)-\sqrt{6} z \sin 2 \theta e^{i (\chi +\phi )}. 
\end{equation}
Similarly, the diagonal enteries for  $M^{\text{diag}}_2$ are
\begin{equation}
  \label{eq:diag eneteries of tm2 mass matrices-11}
  \left( M^{\text{diag}}_{2} \right)_{11} = \frac{1}{2} e^{-2 i \phi } \left(
                                              \sin ^2 \theta  \left(4 b+e^{i
                                              \chi } z\right)-\sqrt{3} z \sin  2
                                              \theta  e^{i (\chi +\phi )}-z \cos
                                              ^2 \theta  e^{i (\chi +2
                                              \phi )}\right),
                                      \end{equation}
                                      \begin{equation}
 \label{eq:diag eneteries of tm2 mass matrices-22}
  \left( M^{\text{diag}}_{2} \right)_{22} = 3 a+e^{i \chi } z, 
\end{equation}
and
\begin{equation}
 \label{eq:diag eneteries of tm2 mass matrices-33}
 \left( M^{\text{diag}}_{2} \right)_{33} = \frac{1}{2} e^{-2 i \phi } \left(\cos ^2 \theta  \left(4 b+e^{i \chi } z\right)+\sqrt{3} z \sin 2 \theta e^{\iota (\chi+\phi)} - z \sin^2 \theta e^{\iota (\chi + 2 \phi)} \right).
\end{equation}%
\begin{figure}[t]
  \centering
 \includegraphics[width=6.29cm]{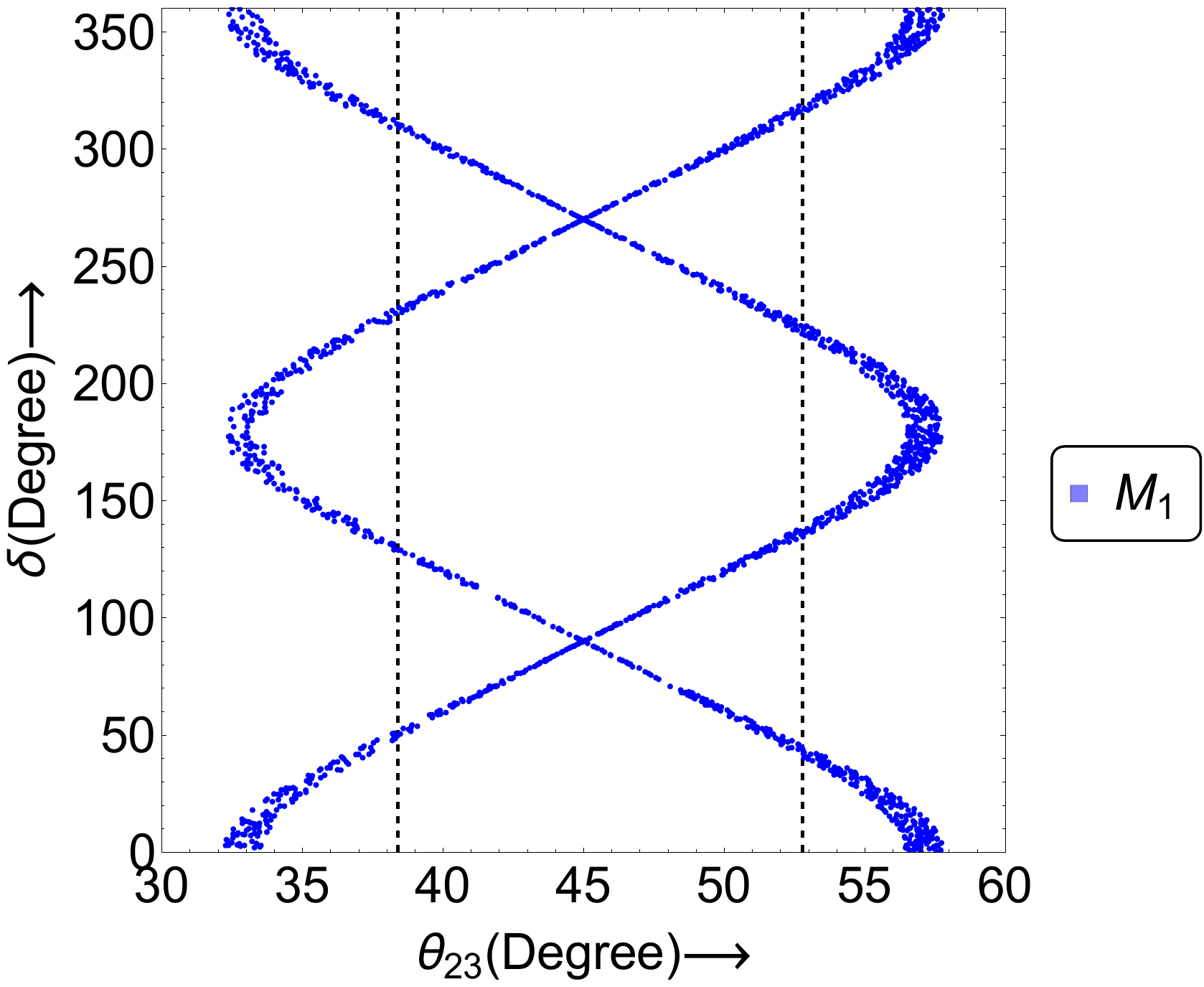} \includegraphics[width=6.29cm]{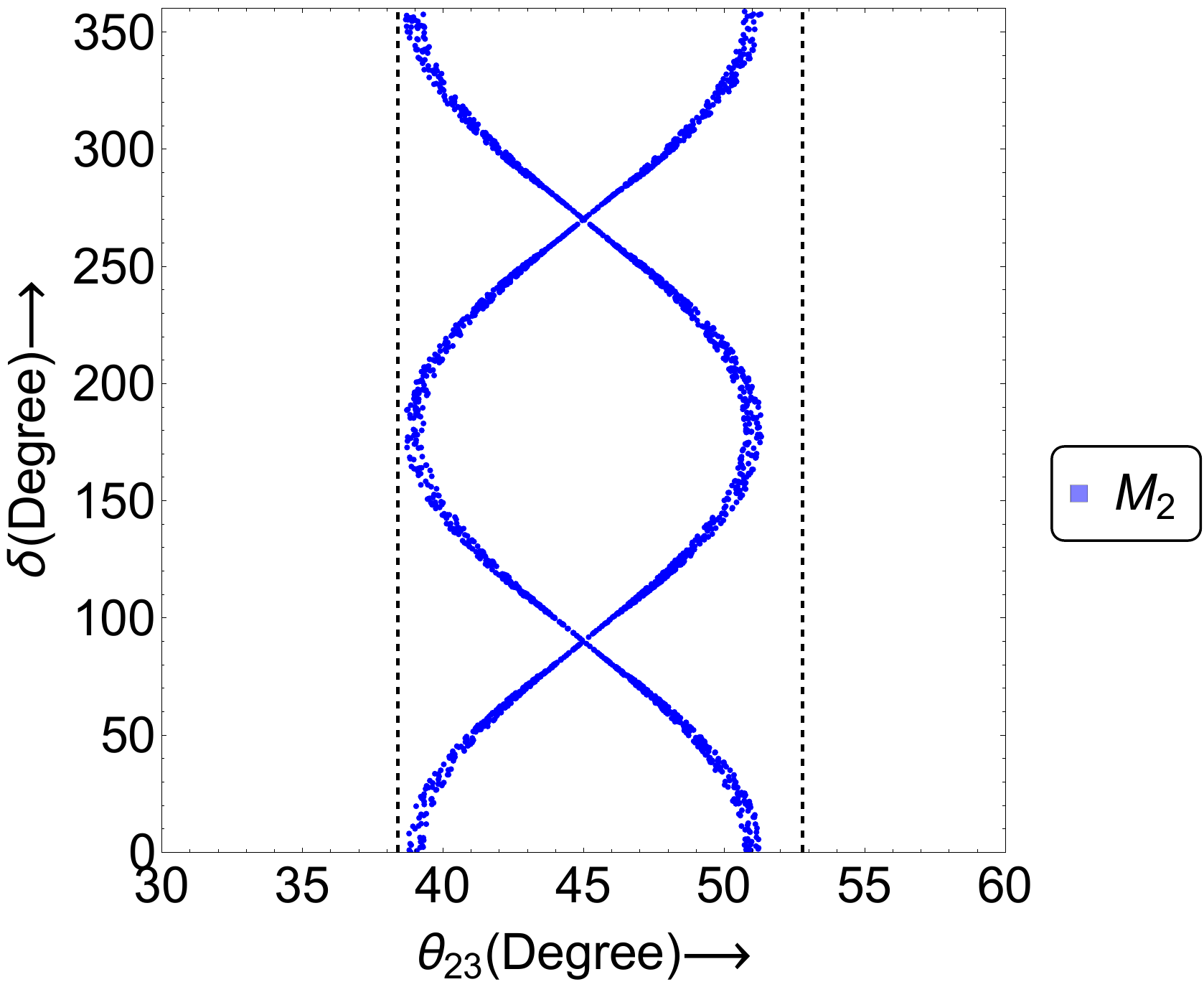}
    \caption{The correlations between atmospheric angle $\theta_{23}$ and CP
      violating phase $\delta$ for both
      the textures M$_1$ and M$_2$. The two vertical dashed lines enclose the
      experimentally allowed region of $\theta_{23}$.}
 \label{fig:th23delta}
\end{figure}
 \begin{figure}[b]
              \centering
          \includegraphics[width=6.29cm]{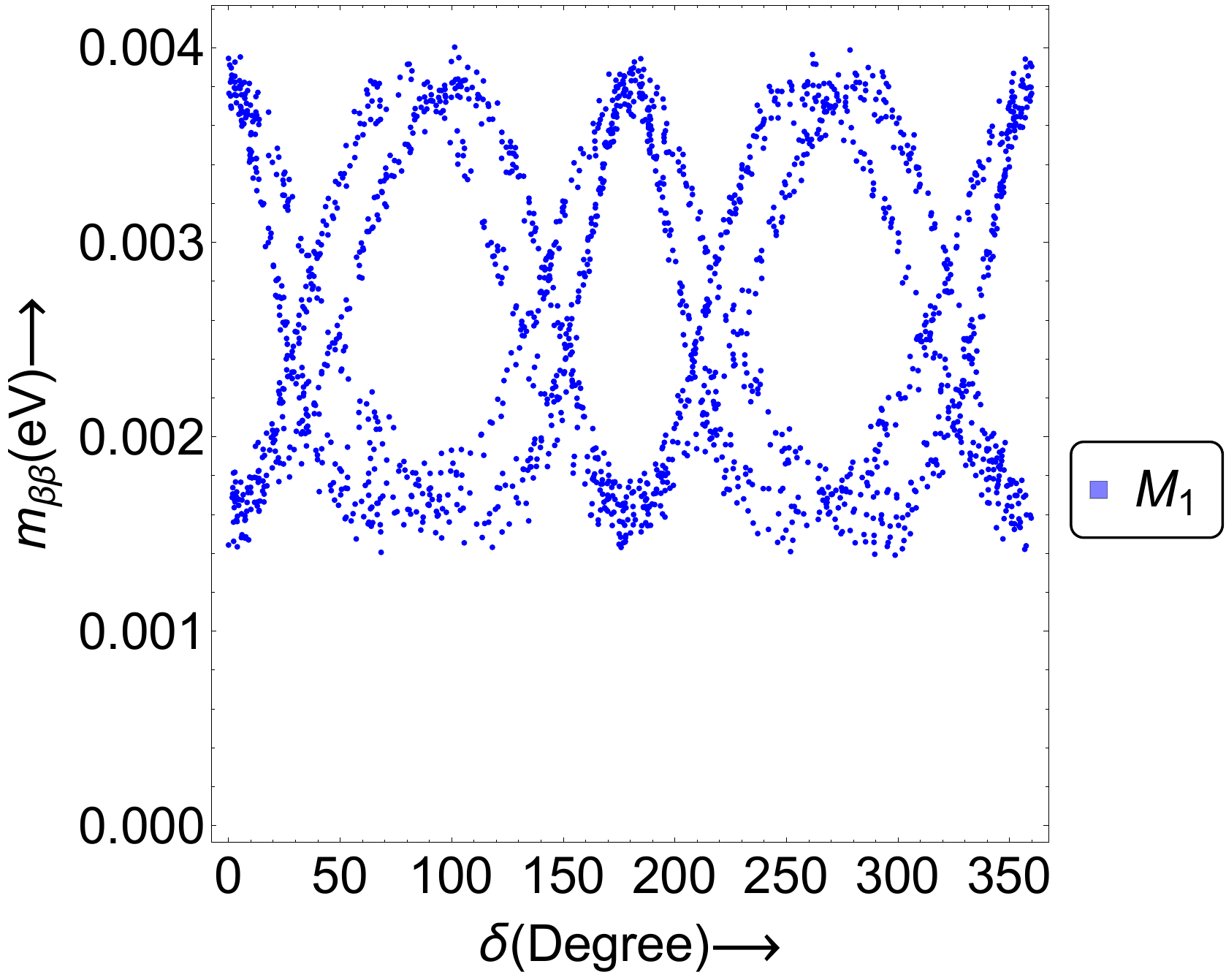} \includegraphics[width=6.29cm]{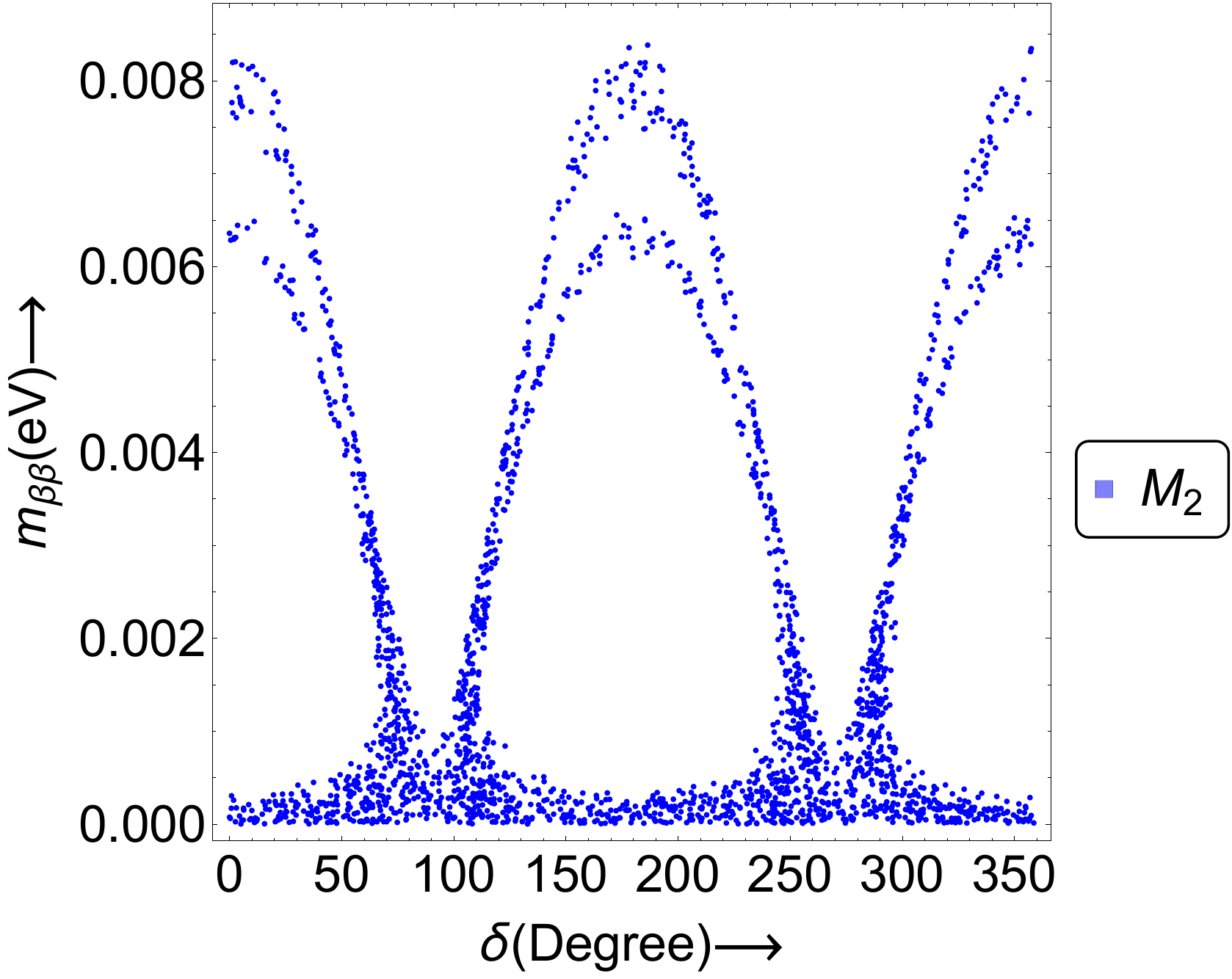}
          \caption{Variation of $m_{\beta \beta}$ with $\delta$ for both the
            textures M$_1$ and M$_2$.\label{fig:mbbdelta}}
\end{figure}
\begin{figure}[t]
    \centering
          \includegraphics[width=6.29cm]{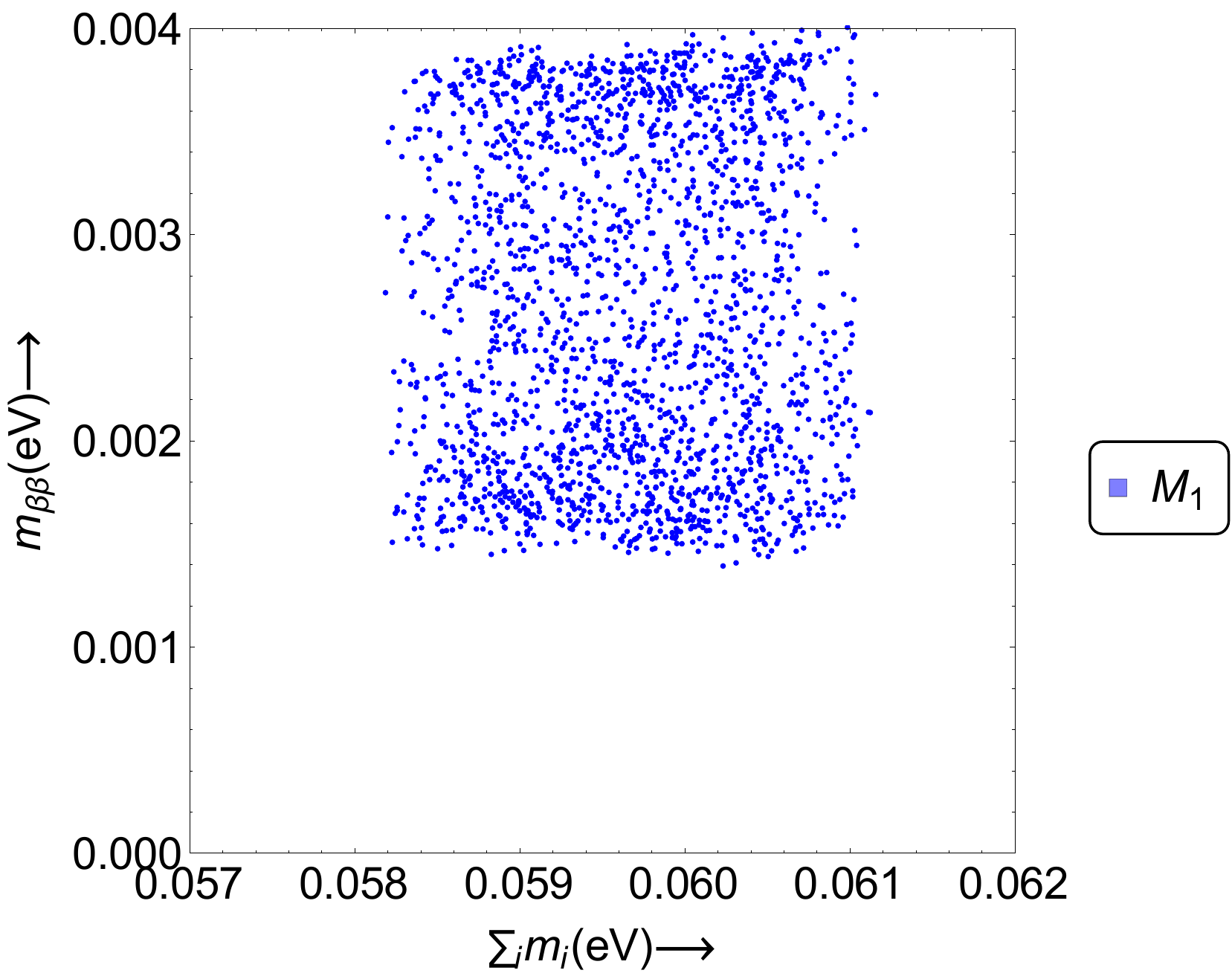} \includegraphics[width=6.29cm]{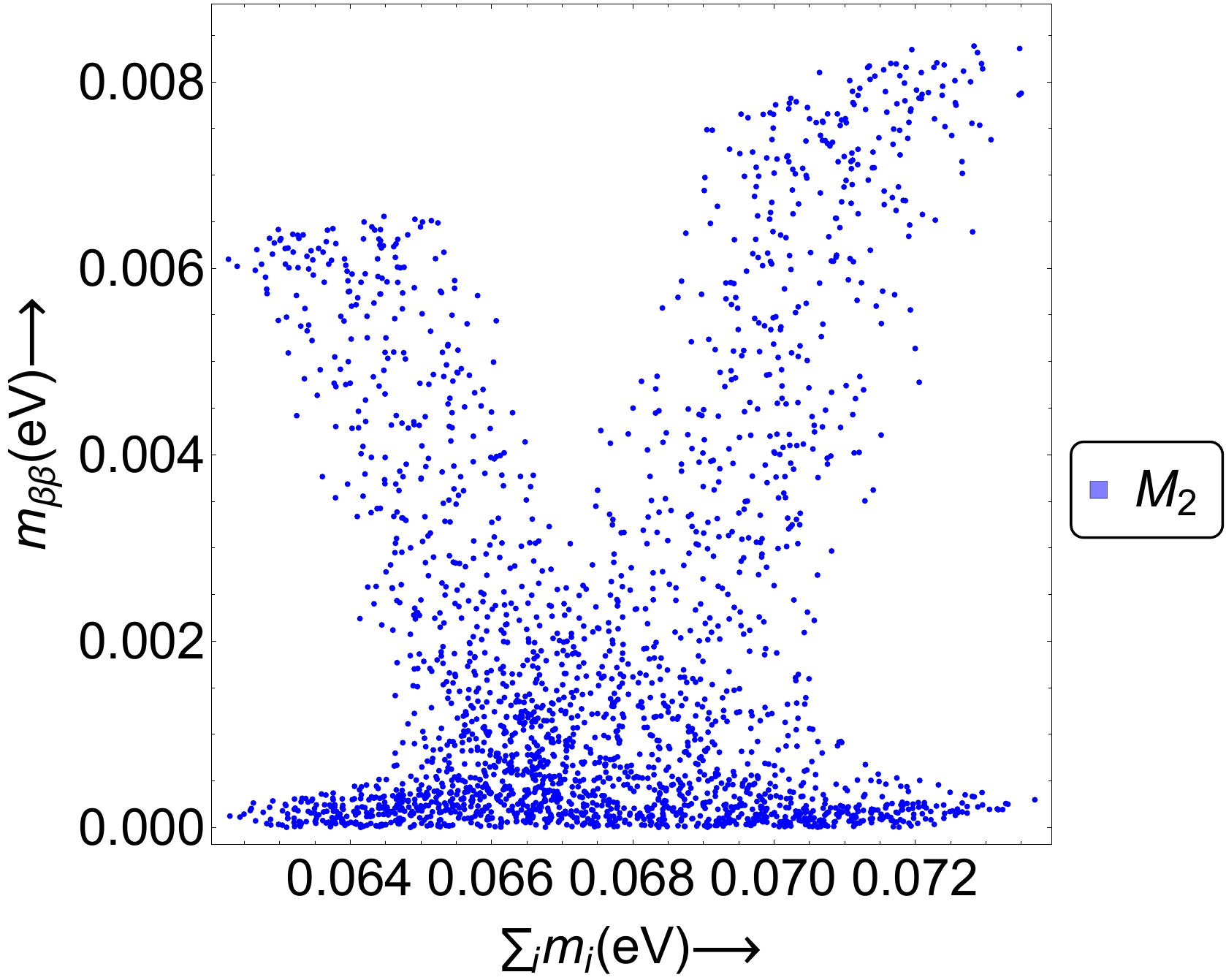}
\caption{The allowed region on ($\sum_{i} m_i$ - $m_{\beta \beta}$) plane for
  textures M$_1$ and M$_2$. \label{fig:mimbb}}
\end{figure}%
Substituting the values of $\theta$ and $\phi$ in these entries, we can
calculate the three neutrino masses for the textures $M_1$ and $M_2$ as
\begin{equation}
  \label{eq:equating of masses with matrix element}
m_1 = \left| \left( M^{\text{diag}}_{i} \right)_{11} \right|, m_2 = \left| \left( M^{\text{diag}}_{i} \right)_{22} \right|, \text{ and } m_3 = \left| \left( M^{\text{diag}}_{i} \right)_{33} \right|
\end{equation}
for $i=1,2$ respectively. We can calculate the mass-squared differences as
$\Delta m^2_{12} = m^2_2 - m^2_1$ and $\Delta m^2_{23} = m^2_3 - m^2_2$ for both
the textures $M_1$ and $M_2$.

We confront these textures with the experimental data \cite{Esteban2016b} by performing a Monte
Carlo analysis of the predictions. We generate a set of N random values for
each of four free
parameters $a_j$, $b_j$, $z_j$, and $\chi_j$ with $j=1,2,3,...N$ uniformaly in
reasonably wide ranges and calculate the predictions for $\theta_{12}$,
$\theta_{23}$, $\Delta m^2_{12}$ and $\Delta m^2_{23}$ using the above
relations. We accept only those points $(a_j , b_j , z_j , \chi_j) $ for which
the calculated values of $\theta_{12}$,
$\theta_{23}$, $\Delta m^2_{12}$ and $\Delta m^2_{23}$ lie in their 3$\sigma$
experimental ranges \cite{Esteban2016b} depicted in Tab.~\ref{tab:experimental}.
We depict $a$, $b$, and $z$ as functions of $\chi$ in Fig.~\ref{massvschi}. The
allowed 3$\sigma$ ranges of $a$,
$b$, and $z$ shown in Tab.~\ref{tab:AllowedRanges}.
\begin{table}[h]
\tbl{Experimental values of the oscillation parameters.}
 {$\begin{array}{cc}
     \toprule
     \text{ Parameters} & \text{ 3$\sigma$ range} \\
          \hline \\
     
     \Delta m^2_{12}/ (10^{-5}) eV^2 &  7.03 \rightarrow 8.09 \\
     
     \Delta m^2_{23}/ (10^{-3}) eV^2 & 2.407 \rightarrow 2.643 \\
     
     \theta_{13}/ ^o & 7.99 \rightarrow 8.90 \\
     
     \theta_{12}/ ^o &  31.38 \rightarrow 35.99 \\

\theta_{23}/ ^o & 38.4 \rightarrow 52.8 \\
     
     \bottomrule
   \end{array}$}
  \label{tab:experimental}
\end{table}

These textures have testable predictions for $\theta_{23}$ and $\delta$. For
texture $M_1$, we find
\begin{eqnarray}
  \label{eq:theta23deltatm1a}
\sin^2 \theta_{23} = \frac{1}{2} \left( 1 + \frac{\sqrt{6} \sin 2 \theta
                         \cos \phi }{3 - \sin^2 \theta
                         } \right), \cot^2 \delta = \cot ^2 \phi  -\frac{6 \sin ^2 2 \theta
  \cot ^2 \phi  }{\left(3-\sin ^2 \theta \right)^2}  ,  \label{eq:theta23deltatm1b}
\end{eqnarray}
and for texture $M_2$, we find
\begin{eqnarray}
  \label{eq:theta23deltatm2}
\sin^2 \theta_{23} = \frac{1}{2} \left(1+ \frac{\sqrt{3} \sin 2 \theta  \cos
                           \phi }{3-2 \sin ^2 \theta }\right), \csc^2 \delta = \csc ^2 \phi -\frac{3 \sin ^2 2 \theta  \cot ^2 \phi
                         }{\left(3-2 \sin ^2 \theta \right)^2},  \label{eq:theta23deltatm2b}
\end{eqnarray}
where $\theta$ and $\phi$ are given by Eqs. (\ref{eq:phifortm1}-\ref{eq:thetafortm1}) and
(\ref{eq:phifortm2}-\ref{eq:thetafortm2}), respectively. The correlation plots
between $\theta_{23}$ and $\delta$ has been depicted in Fig.~\ref{fig:th23delta}. We find
that when $\theta_{23}$ takes its maximal value of
      $45^o$, $\delta$ is either $90^o$ or $270^o$. On the other
      hand, when $\theta_{23}$ takes its extreme values near $40^o$ or $50^o$, $\delta$
      is close to either $0^o$ or $180^o$. We also find that the experimental
      values of $\theta_{23}$ (the region between two vertical dashed lines in
      Fig. \ref{fig:th23delta}) put constraints on $\delta$ and it cannot take
      values around 0 and $\pi$ for the textures M$_1$. The excluded range of
      $\delta$ for the texture $M_1$ is $[130^o , 220^o]$ and $[-30^o,30^o]$. For the texture M$_2$,
      the whole range of $\delta$ is allowed. Such trends can be tested at
      the experiments like T2K \cite{T2K_NON_ZERO_Theta13_PhysRevLett.107.041801} and NO$\nu$A \cite{NOVA_2016_delta_PhysRevLett.116.151806}.

\begin{table}[h]
\tbl{Allowed ranges of the parameters of mass matrix corresponding to the textures
  M$_1$ and M$_2$.}
  {\begin{tabular}{ccc}
    \toprule
    Parameters & \multicolumn{2}{c}{Allowed 3$\sigma$ range} \\
    \colrule
                   & M$_1$ & M$_2$ \\
    \cmidrule(lr){2-2}\cmidrule(lr){3-3} \\
    a & [0.001,0.004]$\cup$[-0.004,-0.001] & [-0.008,0.008] \\
    b & [0.01,0.04]$\cup$[-0.04,-0.01] & [0.021,0.028] $\cup$ [-0.028,-0.021] \\
    z & [0.004,0.006] & [0.009,0.01]\\
    \bottomrule
  \end{tabular}}
    \label{tab:AllowedRanges}
  \end{table}
  
The absolute neutrino masses $m_1$, $m_2$ and $m_3$ can be calculated using
relations given in Eq. (\ref{eq:equating of masses with matrix element}). These masses are not
directly observable at the neutrino experiments. The $\beta$-decay experiments
\cite{Drexlin2013} are sensitive to the effective electron neutrino mass $m_{\beta}$ given as
\begin{equation}
  \label{eq:mb}
m^2_{\beta} = m^2_1 |U_{e1}|^2 + m^2_2 |U_{e2}|^2 + m^2_3 |U_{e3}|^2.
\end{equation}
The neutrino-less double beta decay experiments \cite{Rodejohann2012a} are sensitive to the
effective neutrino mass $m_{\beta \beta} = |(M_{\nu})_{11}|$ given as
\begin{equation}
  \label{eq:mbb}
m_{\beta \beta} = |m_1 U^2_{e1} + m_2 U^2_{e2} + m_3 U^2_{e3}|.
\end{equation}
The cosmological experiments \cite{Couchot2017} are sensitive to the sum of the
neutrino masses $\sum_i m_i = m_1 + m_2 + m_3$. We plot $m_{\beta \beta}$ as
function of $\delta$ in Fig.~\ref{fig:mbbdelta}. We depict a correlation between
$ m_{\beta \beta}$ and $\sum_i m_i$ in Fig.~\ref{fig:mimbb} . We find
  that there is a lower bound on $m_{\beta \beta}$ for textures m$_1$ while
  there is no lower bound on $m_{\beta \beta}$ for texture M$_2$. These predictions
will also be testable in the foreseeable future.
\begin{figure}[h]
   \centering
  \includegraphics[width=6.29cm]{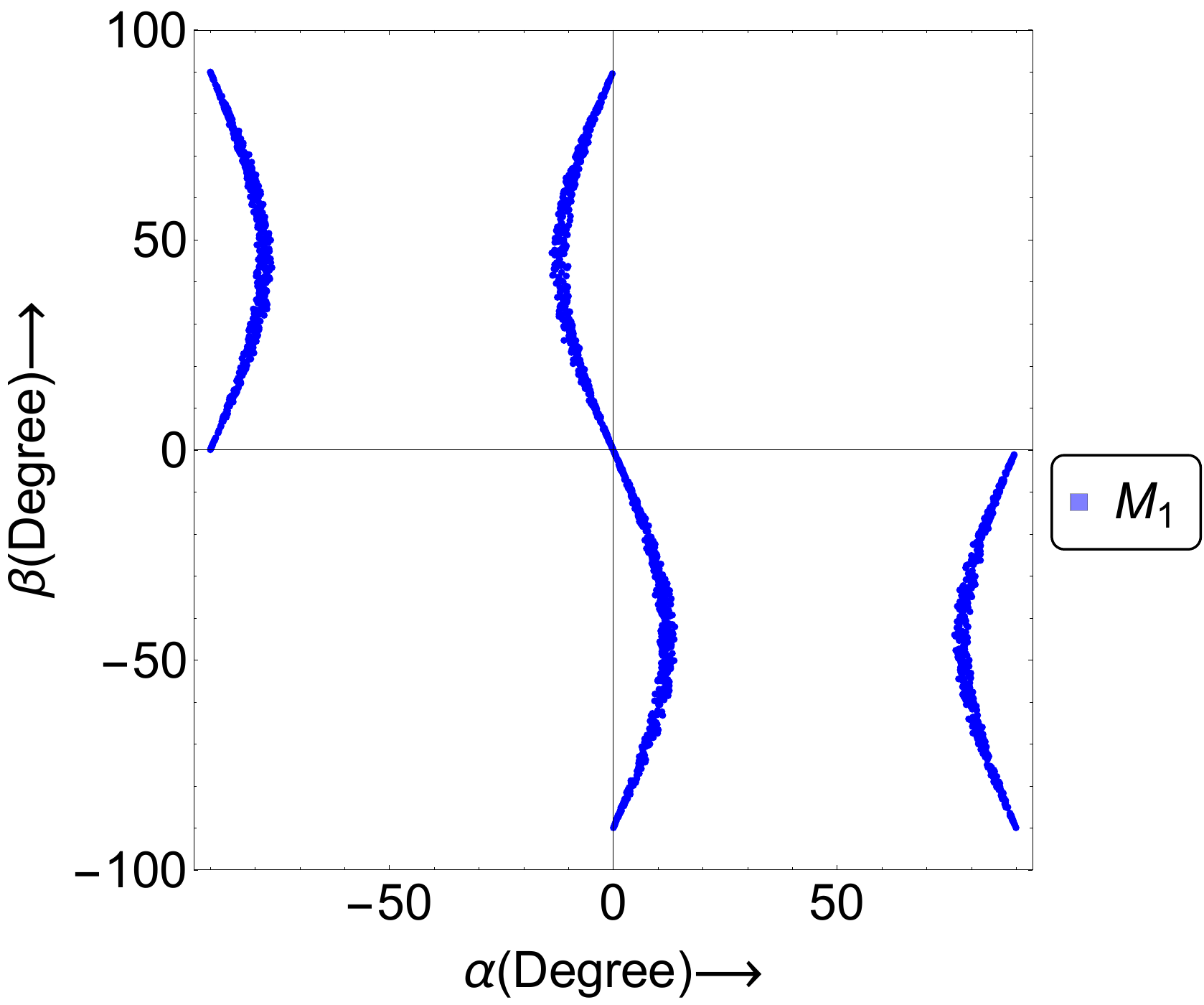}  \includegraphics[width=6.29cm]{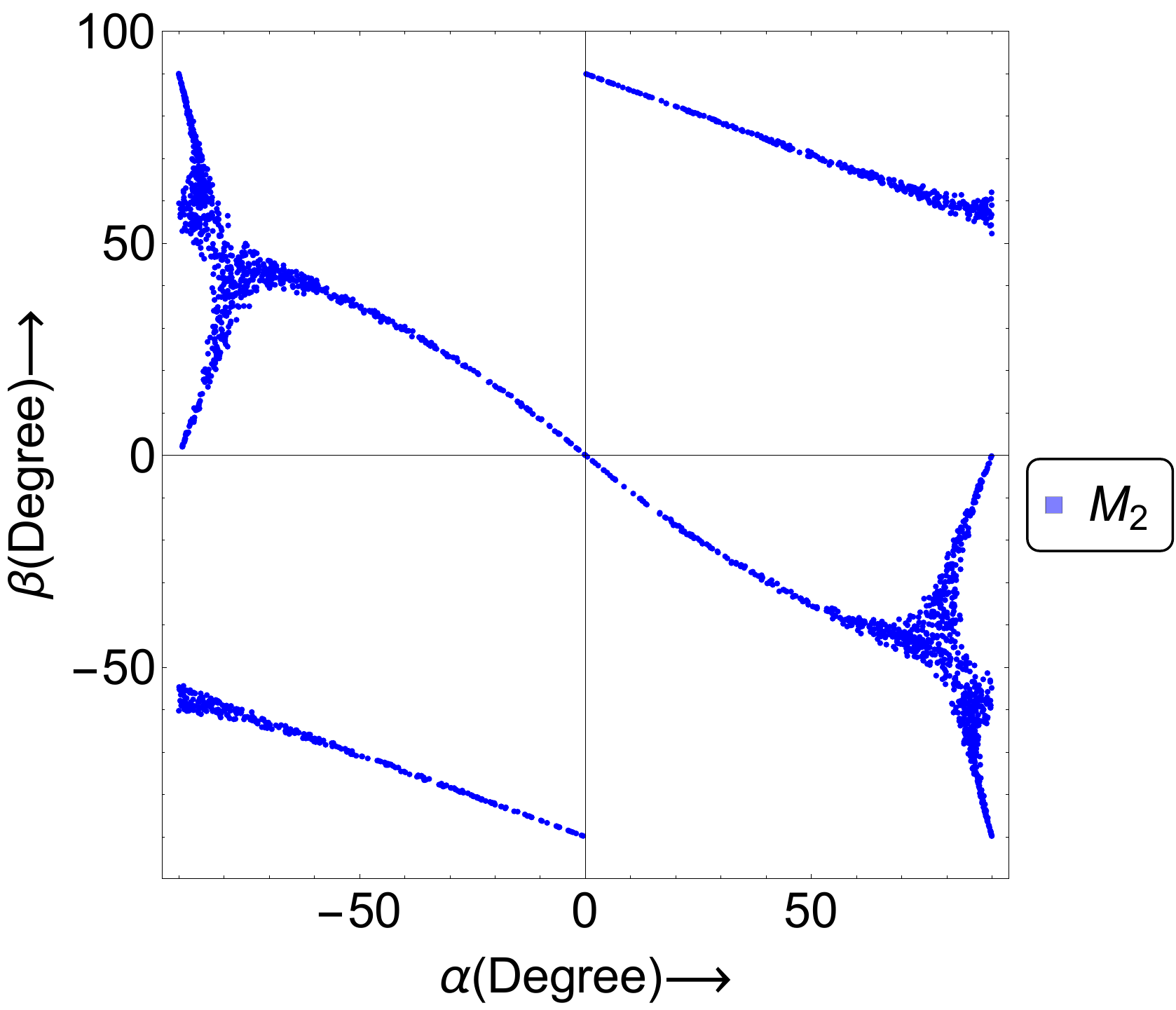}\\
\includegraphics[width=6.29cm]{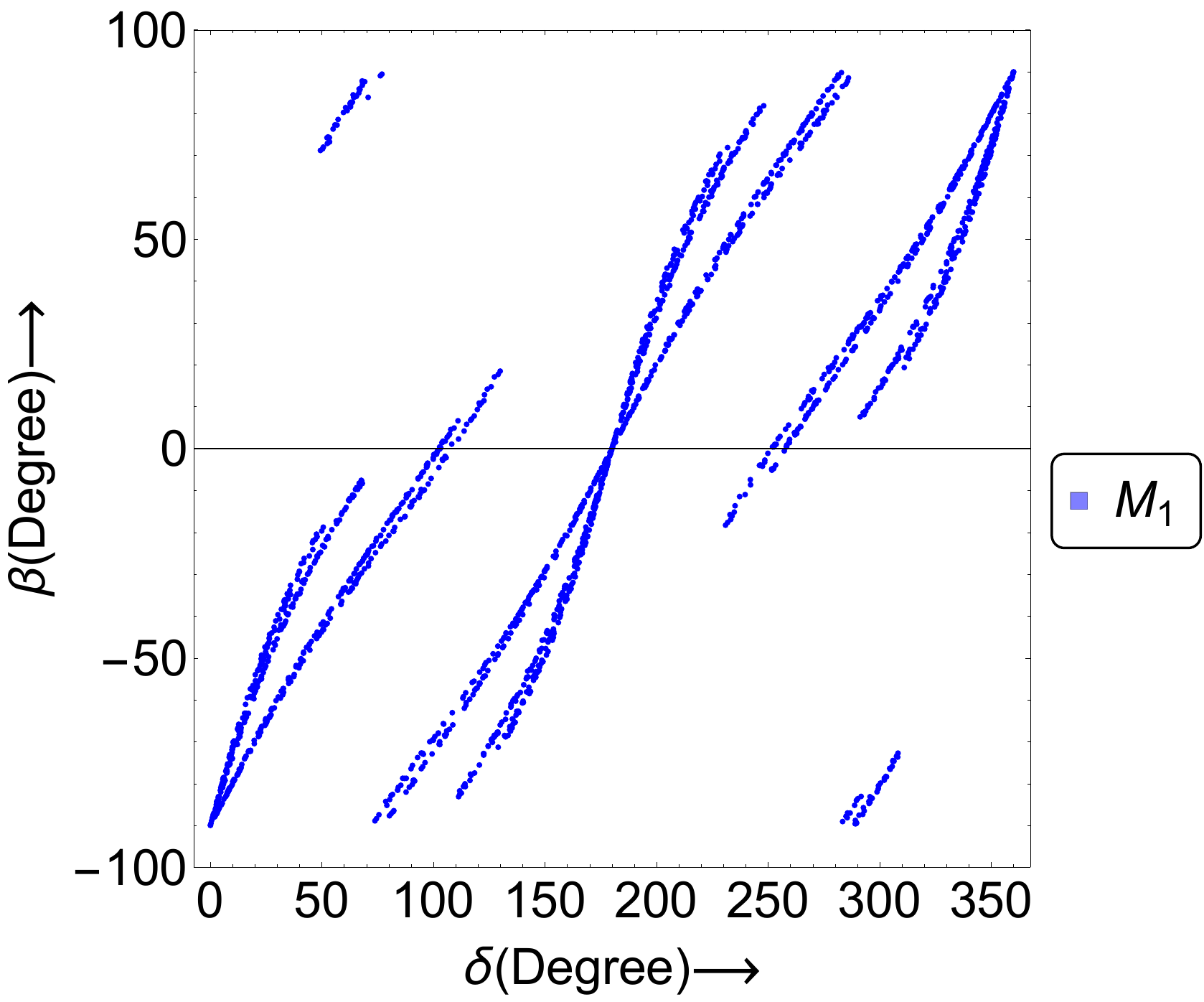} \includegraphics[width=6.29cm]{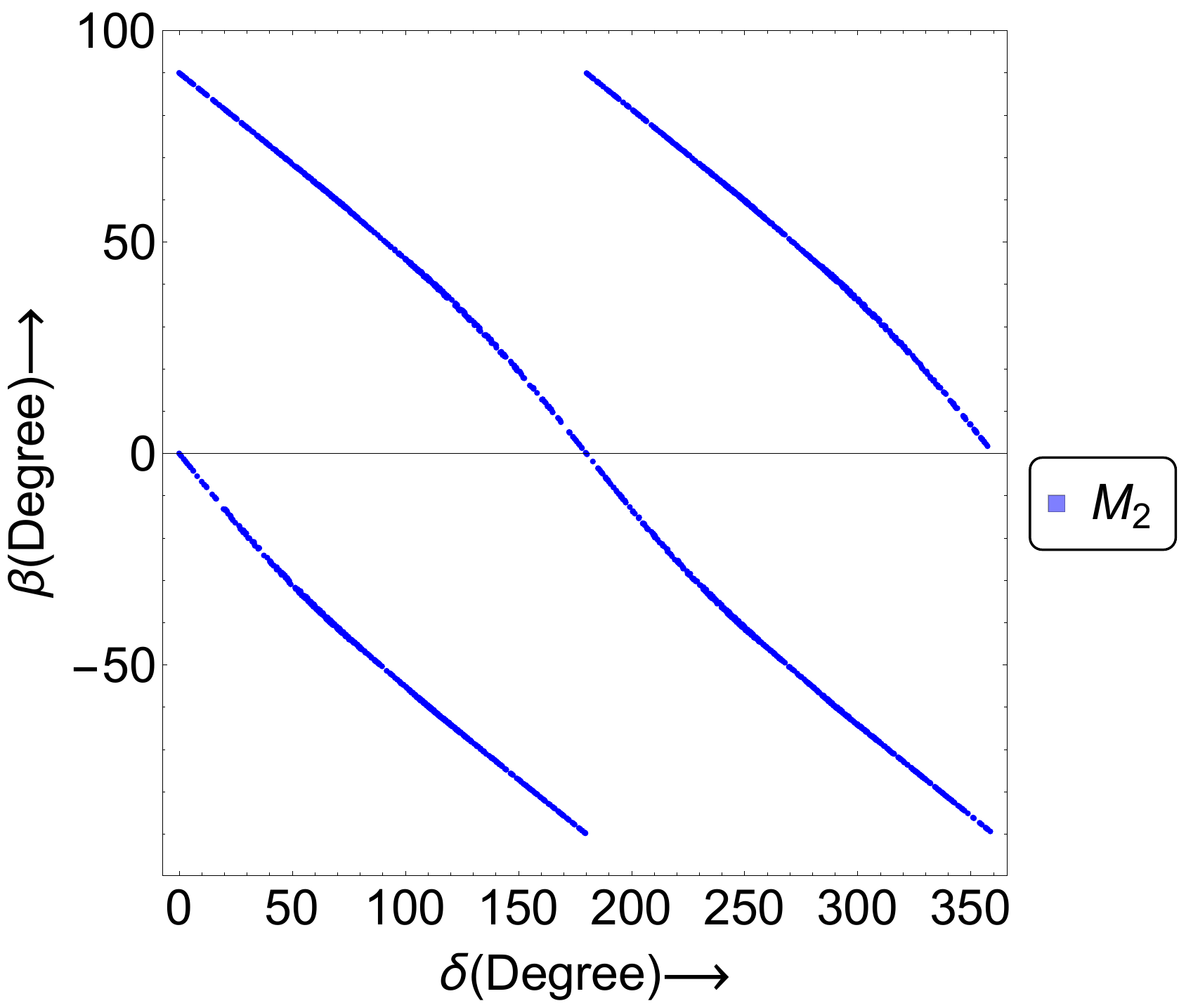}
\caption{Correlation plots for Majorana phases $\alpha$ and $\beta$ for textures
  M$_1$ and M$_2$. }
    \label{fig:majoranaphases}
  \end{figure}
These textures also have interesting predictions for the Majorana phases
$\alpha$ and $\beta$:
\begin{equation}
  \label{eq:majorana phases}
  \alpha = \frac{1}{2} \text{ arg } \left[ \frac{\left( M^{\text{diag}}_{i}\right)_{22}}{\left( M^{\text{diag}}_{i}\right)_{11}} \right] \text{ and }  \beta =  \frac{1}{2} \text{ arg } \left[ \frac{\left( M^{\text{diag}}_{i} \right)_{33}}{\left( M^{\text{diag}}_{i}\right)_{11}} \right]
\end{equation}
for $i=1,2$. We depict the correlation plots for $(\alpha, \beta)$ and
$(\beta,\delta)$ in Fig.~\ref{fig:majoranaphases} for both the textures.

We performed a similar analysis for the textures having inverted hierarchy (
$c=a+2b$ and $d=a+\mu$ in
Eqs. (\ref{eq:mtm1}-\ref{eq:mtm2})). We found that we cannot have values of the
mixing angle $\theta_{13}$ and the ratio $\Delta m^2_{12} / |\Delta
  m^2_{23}|$ in their experimental ranges simultaneously. Hence, these textures
with inverted hierarchy are ruled out. 

In conclusion, we proposed two textures $M_1$ and $M_2$ that correspond to the
TM$_1$ and TM$_2$ mixing schemes respectively and have four free parameters $a$, $b$, $z$, and $\chi$. We show that
these textures are consistent with all the current neutrino data. We find the
allowed ranges for the parameters $a$, $b$, $z$, and $\chi$ and give
predictions for $\theta_{23}$, $\alpha$, $\beta$ and $\delta$. The predictions
for $\theta_{23}$ and $\delta$ are testable at T2K \cite{T2K_NON_ZERO_Theta13_PhysRevLett.107.041801} and
NO$\nu$A\cite{NOVA_2016_delta_PhysRevLett.116.151806}.
\clearpage
\bibliographystyle{ws-mpla}
\bibliography{library,first}
\end{document}